\documentclass{aa}
\usepackage{graphicx}
\usepackage{txfonts}
\def\erg{${\rm erg ~cm^{-2} ~s^{-1}}$ }
\def\sex{{\tt SExtractor}}
\def\hel{{\tt Hellas2XMM}}
\def\ks{$K_s$}
\def\rk{$R$--$K$}
\def\xmm{{XMM--{\it Newton}}}
\def\zmin{$z_{\rm min}$}
\def\gsimeq{\hbox{\raise0.5ex\hbox{$>\lower1.06ex\hbox{$\kern-1.07em{\sim}$}$}}}
\def\lsimeq{\hbox{\raise0.5ex\hbox{$<\lower1.06ex\hbox{$\kern-1.07em{\sim}$}$}}}
\begin{document}
%
%
  \title{The HELLAS2XMM survey}

  \subtitle{V. Near-Infrared observations of X-ray sources with \emph{extreme}
            X/O ratios.
   \thanks{based on observations collected at the European Southern
    Observatory, Paranal, Chile (ESO Programme 70.A-0657) and La Silla,
    Chile (ESO Programme IDs: 66.A-0520, 67.A-0401, 68.A-0514).
    Based also on observations made with the \xmm,
    an ESA science mission with instruments and contributions directly
    funded by ESA member states and the USA (NASA).}}

  \author{M.~Mignoli\inst{1}
          \and
          L.~Pozzetti\inst{1}
          \and
          A.~Comastri\inst{1}
          \and
          M.~Brusa\inst{1,2}
          \and
          P.~Ciliegi\inst{1}
          \and
          F.~Cocchia\inst{3}
          \and
          F.~Fiore\inst{3}
          \and
          F.~La~Franca\inst{4}
          \and
          R.~Maiolino\inst{5}
          \and
          G.~Matt\inst{4}
          \and
          S.~Molendi\inst{6}
          \and
          G.C.~Perola\inst{4}
          \and
	  S.~Puccetti\inst{3}
          \and
          P.~Severgnini\inst{7}
          \and
          C.~Vignali\inst{1}
          }

   \offprints{M. Mignoli}
   \authorrunning{M. Mignoli et~al.}
   \titlerunning{The {\it HELLAS2XMM} Survey V.}

   \institute{INAF -- Osservatorio Astronomico di Bologna, 
    via Ranzani,~1 I-40127 Bologna, Italy\\
              \email{mignoli,lucia,comastri,ciliegi,l\_vignali@bo.astro.it}
         \and
   Dipartimento di Astronomia Universit\`a di Bologna,
    via Ranzani 1, I--40127 Bologna, Italy\\
              \email{brusa@bo.astro.it}
         \and
   INAF -- Osservatorio Astronomico di Roma,
    via Frascati 33, I--00040 Monteporzio, Italy\\
              \email{cocchia,fiore,puccetti@mporzio.astro.it}
         \and
   Dipartimento di Fisica Universit\`a di Roma Tre,
    via Vasca Navale 84, I--00146 Roma, Italy\\
              \email{lafranca,matt,perola@fis.uniroma3.it}
         \and
   INAF -- Osservatorio Astrofisico di Arcetri,
    Largo E. Fermi 5, I--50125, Firenze, Italy\\
              \email{maiolino@arcetri.astro.it}
         \and
   IASF -- CNR, Istituto di Fisica Cosmica,
    via Bassini 15, I--20133, Milano,  Italy\\
              \email{silvano@mi.iasf.cnr.it}
         \and
   INAF -- Osservatorio Astronomico di Brera, 
    via Brera 28, I--20121 Milano, Italy\\
              \email{paola@brera.mi.astro.it}
             }

   \date{Received .....; accepted .....}

   \abstract{We present the results of deep near-infrared observations
   (with ISAAC at VLT) of eleven hard X-ray selected sources in the \hel \
   survey, with faint optical magnitude ($R\,\gsimeq\,$24)
   and high X-ray-to-optical flux ratio.
   All but one of the sources have been detected in the \ks \ band, 
   with bright counterparts (\ks$<$19) and very red colors ($R-K>5$), 
   and therefore belong to the ERO population.
   The quality of the near-infrared data is such that we can
   take advantage of the sub-arcsec seeing to obtain
   accurate morphological information. A detailed analysis of the
   surface brightness profiles allows us to classify all of the
   near-infrared counterparts. There are two point-like objects, seven
   elliptical (bulge) galaxies and one source with an exponential profile.
   None of the extended sources shows any evidence for the presence
   of a central unresolved object tracing the putative X-ray emitting AGN.
   Using both the \rk \ colors and the morphological information,
   we have estimated  for all the sources a ``minimum photometric redshift'',
   ranging between 0.8 and 2.4; the elliptical hosts
   have $z_{\rm min}=0.9-1.4$. We computed the X-ray properties
   using these redshifts: most of the sources have
   N$_H>10^{22}$ cm$^{-2}$, with unabsorbed
   X-ray luminosities up to 10$^{45}$~erg~s$^{-1}$
   in the intrinsic 2$-$10~keV band.
   These objects therefore belong to the long-sought population of obscured
   (Type~II) quasars and, from a statistical point of view,
   they turn out to be a non-negligible
   fraction (about 10\%) of the most luminous AGN.
   Selecting the high X/O sources for a follow-up study in the near-infrared 
   is therefore a powerful technique aimed at studying at high redshift
   the hosts of Type~II AGN, whose obscured nuclei do not affect the
   host galaxy morphologies. Overall, our results seem to indicate that the
   hosts are mostly elliptical galaxies at $z\sim1$, and that these near-IR
   bright objects would be among the most massive spheroids at these
   epochs.
   }

   \maketitle

%
%
\section{Introduction}
%
\par\noindent
With the advent of imaging X-ray telescopes operating 
onboard {\it Chandra} and \xmm, a large fraction 
($>80$\%, Alexander et~al. \cite{alex03}; Giacconi et~al. \cite{giacco02};
Hasinger et~al. \cite{hasing01}; Mushotzky et~al. \cite{mush})
of the Cosmic X-ray Background (XRB) in
the $2-8$ keV band has been resolved into discrete sources.
The detailed study of the nature of the hard X-ray source population
is pursued combining deep ($1-2$~Ms) {\it Chandra} pencil beam observations 
in the CDF-N and CDF-S, with shallower large area surveys (i.e. 
{\tt \hel } -- Baldi et~al. \cite{paper1}; {\tt Champ} -- Green et~al.
\cite{champ}; {\tt SEXSI} -- Harrison et~al. \cite{sexsi}).
There are several issues which can be properly addressed only 
by surveying several square degrees at relatively bright fluxes, namely:  
i) the search for rare objects, which requires sufficiently wide areas to
be discovered, ii) a more uniform coverage of the Hubble diagram and
iii) a detailed study --- by means of X-ray spectroscopy
and/or by multi-wavelength follow-up --- of X-ray sources 
which in the deep surveys are too faint for a similar approach.

In this respect we note that an interesting new population of X-ray 
sources is present in both the deep and shallow surveys.
These objects are characterized by values of their
X-ray--to--optical flux ratio $f_{X-ray}/f_{opt}>10$
(hereinafter X/O; see Maccacaro et~al. \cite{maccacaro} for a definition),
which are significantly larger than those observed for
soft X-ray selected AGN in the ROSAT surveys (0.1$<$ X/O $<$10;
Lehmann et~al. \cite{lehmann}; Zamorani et~al. \cite{zamorani}). 
Almost by definition, sources with X/O $>10$ have faint
optical magnitudes. This is even more true for the faint X-ray sources 
discovered in the {\it Chandra} deep fields. For example,
a 2--8 keV flux of 10$^{-15}$(10$^{-16}$) \erg \ and an X/O ratio
$\sim$10 correspond to $R$ magnitudes $\sim$25.5 ($\sim$28),
challenging (well beyond) the spectroscopic capabilities of the
10m-class telescopes.
 
On the other hand, at the brighter fluxes covered by the \hel \
survey ($1-40\times 10^{-14}$ \erg), the magnitudes of the
optical counterparts of high X/O sources are of the order of
$R \simeq 24$ or brighter, making spectroscopic follow--up
observations feasible. Indeed, more than half of the \hel \ sources
with X/O $>10$, for which good quality VLT spectra are available,
are classified as Type~II QSOs 
on the basis of the lack of broad optical lines and high X-ray
luminosity ($L_{2-10 keV} > 10^{44}$~erg~s$^{-1}$; see Fiore et~al.
\cite{H2X1dF}).
It seems reasonable to argue that most of the sources characterized 
by a high value of X/O are high redshift, obscured AGN.
If this were the case, they could contribute to reducing the disagreement
between the redshift distribution predicted by XRB synthesis models
and that observed in deep {\it Chandra} and \xmm \ fields  
(Hasinger \cite{hasing03}; Gilli \cite{gilli}), and provide an important
contribution to the total energy density of the background light.
Furthermore, sources with even more extreme values of X/O ($\gg$10) are
present in the \hel \ survey. These objects are again very challenging
to be followed-up with the optical spectrographs attached to
10-m class telescopes, as they are both very faint ($R\;\gsimeq\;25$) 
and spread over a large area of sky, making unfeasible any deep
exposures with multi-object instruments. 
An alternative approach is to make use of the observed property
that the optically fainter X-ray selected sources have, on average,
redder colours (Giacconi et~al. \cite{giacco01}; Lehmann et~al.
\cite{lehmann}; Alexander et~al. \cite{alex02}).
This motivates the choice of the near infrared as the natural band to
carry out a follow-up study of high X/O sources.

\begin{figure}
   \centering
   \includegraphics[width=0.49\textwidth]{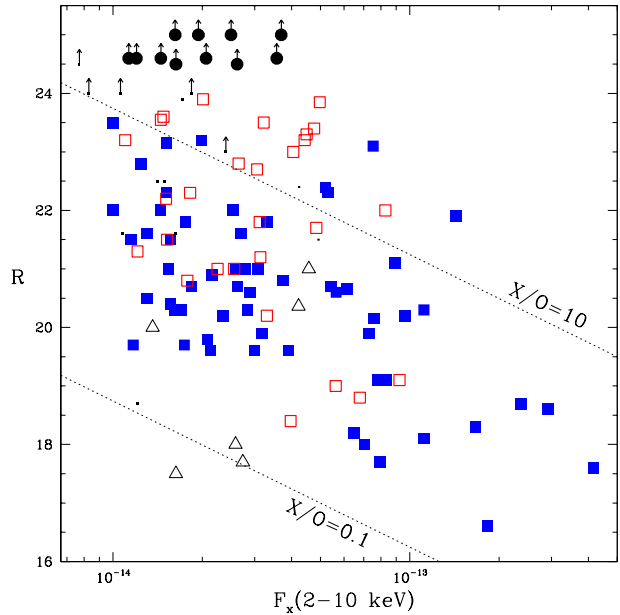}
   \caption{The optical ($R$~band) magnitude versus the X-ray (2-10 keV) flux
   for the \hel \ 1dF sample. Different symbols identify different source
   classes: filled squares = broad line AGN; empty squares = narrow line AGN;
   triangles = early-type galaxies; small dot = spectroscopically unidentified 
   sources.
   The selected sources observed with $ISAAC$ are marked with large
   dots. The dotted lines indicate the loci of constant X-ray-to-optical
   flux ratio, delimiting the region of ~``conventional'' AGN.
   }
    \label{fig_xo_R}
    \end{figure}

On the basis of these considerations, we started a pilot study
in the \ks\ band, selecting a restricted but meaningful sample
of \hel \ sources that satisfy a well defined selection criterion:
bright X-ray fluxes \hbox{($F$(2$-$10~keV) $> 10^{-14}$ \erg)}, high
X-ray--to--optical flux ratios (X/O $> 10$) and no detection in the $R$~band
(see Fig.\ref{fig_xo_R}).

In Section 2 we present the target selection. The infrared data analysis 
and properties of the counterparts are discussed in Sections 3 and 4,
respectively. 
The results are presented in Section 5 and summarized in Section 6.
A concordance cosmological model, with $H_0 = 70$ Km s$^{-1}$ Mpc$^{-1}$,
$\Omega_m$ = 0.3, $\Omega_\Lambda$ = 0.7, is assumed throughout
the paper.

%
%
\section{The \hel \ Survey and the Selection of the X-ray Targets}
%
%
\begin{table*}
       \caption{X-ray properties of the selected sources}
       \label{tabsam}
       \vspace{0.2cm}             
\begin{minipage}{0.99\textwidth}
\begin{tabular}{lllcl}
\hline\hline\noalign{\smallskip}
SOURCE ID. & RA & DEC & 2-10 keV flux & log(X/O) \\
  & (J2000) & (J2000) &(10$^{-14}$\erg) &  \\
\noalign{\smallskip}\hline\noalign{\smallskip}
Abell2690~~~\#075 & 23 59 56.6 & $-$25 10 20 & 2.1$\pm$ 0.7  & $>1.65$ \\
Abell2690~~~\#029 & 00 01 11.6 & $-$25 12 03 & 3.6$\pm$ 0.9  & $>1.89$ \\
BPM16274~~\#069   & 00 50 30.7 & $-$52 00 46 & 2.6$\pm$ 0.5  & $>1.72$ \\
BPM16274~~\#181   & 00 50 31.6 & $-$52 06 30 & 1.6$\pm$ 0.3  & $>1.50$ \\
PKS0312-77~~\#045 & 03 10 19.0 & $-$76 59 58 & 1.9$\pm$ 0.6  & $>1.79$ \\
PKS0312-77~~\#031 & 03 11 13.9 & $-$76 53 59 & 1.2$\pm$ 0.3  & $>1.42$ \\
PKS0312-77~~\#036 & 03 13 43.5 & $-$76 54 26 & 1.5$\pm$ 0.5  & $>1.50$ \\
PKS0537-28~~\#111 & 05 39 11.5 & $-$28 37 18 & 3.7$\pm$ 0.6  & $>2.19$ \\
PKS0537-28~~\#054 & 05 39 45.3 & $-$28 49 11 & 1.6$\pm$ 0.5  & $>1.71$ \\
PKS0537-28~~\#091 & 05 40 21.2 & $-$28 50 38 & 2.5$\pm$ 0.7  & $>2.06$ \\
PKS0537-28~~\#037 & 05 41 00.4 & $-$28 39 05 & 4.9$\pm$ 1.3  & $>1.80$ \\
\noalign{\smallskip}\hline
\end{tabular}
\end{minipage}
\end{table*}
%
The \hel \ Survey is a program of multiwavelength 
follow--up observations of hard X-ray selected sources 
serendipitously discovered in \xmm \ fields
over $\sim 4$ deg$^{2}$. 
All the details on X-ray data analysis and 
source detection procedure have been already discussed in Baldi 
et~al. (\cite{paper1}).
To date, we have performed a complete optical follow--up program in five 
\xmm \ fields covering a total of about 0.9 square degrees 
(the \hel-1dF, Fiore et~al. \cite{H2X1dF}).
For all of the 122 sources in the \hel-1dF sample
we obtained relatively deep ($R=24-25$) imaging 
using the ESO~3.6m and the TNG telescopes; 
106 sources (87\%) were found to be associated with optical
counterparts brighter than $R=24$. 
For 97 objects 
redshift and classification have been obtained using ESO spectrographs 
(EFOSC2@3.6m and FORS1@VLT/UT1). 
The source breakdown is summarized in Fig.\ref{fig_xo_R} 
(see Brusa et~al. \cite{p0312} 
and Fiore et~al. \cite{H2X1dF} for all the details concerning the imaging 
and spectroscopic observations and the optical identification process).

The majority of the unidentified sources 
(upward arrows in Fig.\ref{fig_xo_R})  
are optically faint and are characterized by X/O $>10$.
The eleven X-ray brightest sources undetected in the optical
with $R>$24.5, were selected to be observed in the \ks \ band
(shown in Fig.\ref{fig_xo_R}
as large dots in the upper part of the diagram).
Most of them have a hard X-ray spectrum (see Sect. 5.3), 
lending further support to the hypothesis 
that they are obscured at both X-ray and optical wavelengths. 
The basic properties of the selected sample are given in Table~\ref{tabsam}.
%
%
\section{Observations and Data Analysis}
%
   \begin{figure*}
   \centering
   \includegraphics[width=0.99\textwidth]{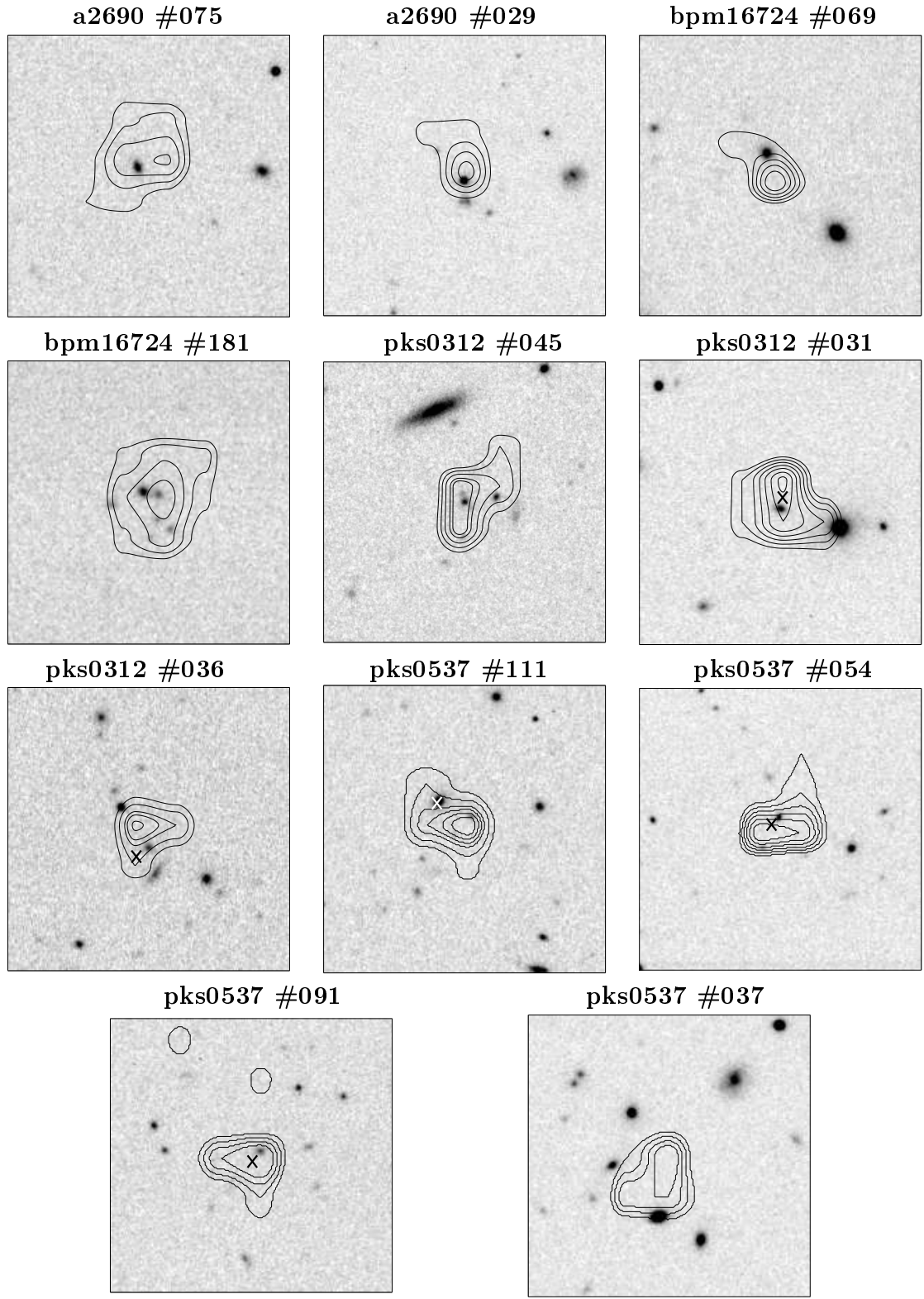}
   \caption{ISAAC \ks \ images, centered on the X-ray source centroid
   and 30\arcsec \ wide. 
   North is to the top and east to the left.
   The X-ray contours are overlaid on each image.
   For 5 objects in the {\tt PKS0312$-$77} and {\tt PKS0537$-$28} fields we
   also show as a small cross the position of the source detected by
   {\it Chandra}.}
    \label{fig_fc}
    \end{figure*}
%
 \subsection{Near-Infrared Observations}
 \par\noindent
 Deep \ks\footnote{The \ks \ ({\it K short}) filter is centered at shorter
 wavelength than the standard K filter in order to reduce thermal background.}
 observations were obtained with the Infrared
 Spectrometer And Array Camera ($ISAAC$, see Moorwood et~al. \cite{isaac})
 mounted on the ESO VLT-UT1 ({\it Antu}) telescope. $ISAAC$ is equipped with
 a Rockwell Hawaii 1024$\times$1024 HgCdTe array, with a pixel scale of
 0.147\arcsec/pixel and a field of view of 2\farcm5 $\times$ 2\farcm5.
 The observations were collected in service mode over several nights
 in September 2002 under good seeing conditions ($<$0\farcs8).
 All the 11 fields were imaged with the same observing strategy, using
 the standard ``auto-jitter'' mode with the telescope being randomly offset
 by amounts of up to 30\arcsec \ between individual short exposures for optimal
 background subtraction. Individual exposures have integration times of 
 6$\times$10 sec (n.sub-integrations $\times$ detector integration times),
 and each X-ray source field was imaged with 36 such jittered
 individual exposures for a common total integration time of 2160~sec. 
 The ESO calibration plan for $ISAAC$ provides the necessary calibrations for
 our purposes, including detector darks, twilight flats and nightly photometric
 zero points, which were obtained by observing near-IR standard stars from
 the list of Persson et~al. (\cite{persson}).
 
 The data reduction was performed in two stages: individual 
 raw frames were first corrected for bias and dark current by subtracting
 a median dark frame, and flat-fielded using an average differential
 sky flat-field image. For the sky subtraction and
 image co-adding we used {\tt DIMSUM}\footnote{Deep Infrared Mosaicing Software,
 developed by P.~Eisenhardt, M.~Dickinson, A.~Stanford and J.Ward.},
 a contributed  package of IRAF\footnote{IRAF is distributed by the
 National Optical Astronomy Observatories, which is operated by the Association
 of Universities for Research in Astronomy, Inc, under contract
 with the National Science Foundation.}. To subtract the background,
 {\tt DIMSUM} follows a two-step procedure. In the first step,
 a median sky is computed for each individual image from the six adjacent
 frames. The shifts between the sky-subtracted images are then computed
 and all of the frames are stacked to produce a ``raw'' co-added image.
 This image is used to build an object mask, which is used to flag
 the pixels belonging to sources in the individual frames. 
 The second step repeats the operations of the first one, but
 includes the object masking and a rejection algorithm to remove cosmic rays.
 The final co-added images show flat, zero-leveled backgrounds and
 stable PSFs over the field of view. 
 In Table~\ref{tabobsir} we summarize the relevant information of the
 \ks \ observations.
  
 Object detection was carried out running \sex \ 
 (Bertin \& Arnouts \cite{bertin}) on the final co-added
 images. Thanks to the high uniformity of the observations 
 (see Table~\ref{tabobsir}), we could
 use the same detection criterion for all of the data: we first convolved the
 data with a 0\farcs6-diameter Gaussian kernel and then classified as
 detected objects those with at least 5 connected pixels above 2.5 times the
 sky noise. As a measure of the \emph{total} source magnitude we adopted the 
 Kron-like magnitude, calculated by \sex \ by fitting elliptical `adaptive'
 apertures to each detection.
 The magnitudes are given in the Vega system. 
 The star/galaxy classification was performed using a plot of the 
 half-flux radius versus the \ks \ magnitude. 
 Point-like objects are easily identified down to  \ks $\simeq$ 19.0, 
 while the fainter objects were assumed
 to be galaxies: at these faint magnitudes and at high Galactic latitudes,
 the stellar contamination is negligible.
 The final catalog contains $\approx$1800 
 galaxies up to \ks=21.5 in the area
 spanned by the $ISAAC$ observations of $\approx 70$ arcmin$^2$ 
 (the edges of the mosaics were discarded because of the low signal-to-noise
 ratio).  The galaxy counts agree well with the
 K20 survey counts (Cimatti et~al. \cite{K20_3}), and the comparison with
 deeper surveys (Saracco et~al. \cite{saracco}, Bershady,  Lowenthal \& Koo
 \cite{bershady}, Moustakas et~al. \cite{mous}) suggests that we reach a
 50\%~completeness at \ks$\approx$21. 
 The known discrepancy in the near-infrared surface density of galaxies
 between different surveys, to be mainly ascribed to cosmic variance, 
 is also present in our data, where we find a factor
 up to $\sim$1.8 between the surface densities at \ks=20 in different fields.
 This result is not at variance with the above statement of high uniformity
 in our data, since we must also take into account the nature of our targets,
 probably obscured and powerful AGN, and the known overdensity
 of near-infrared sources around high-z quasars (Hall \& Green \cite{hall}).
 Finally, the incompleteness at faint magnitudes is not a drawback
 for this study, since all of the counterparts of our hard X-ray selected
 sources are bright in the near infrared, with \ks$\ll 20$ (see Sect. 4.1).
 
\begin{table}
       \caption{\ks \ Observations Characteristics}
       \label{tabobsir}
       \vspace{0.2cm}
\begin{minipage}{0.99\textwidth}
\begin{tabular}{llll}
\hline\hline\noalign{\smallskip}
SOURCE ID. & AIRMASS & SEEING\footnote{FWHM in arcsec.} & 
$\sigma_{\mu}(sky)$\footnote{1$\sigma$ sky fluctuations in mag/arcsec$^2$.} \\
\noalign{\smallskip}\hline\noalign{\smallskip}
Abell2690~~~\#075 & 1.39 & 0.58 & 22.09 \\
Abell2690~~~\#029 & 1.17 & 0.49 & 21.92 \\
BPM16274~~\#069   & 1.16 & 0.76 & 22.17 \\
BPM16274~~\#181   & 1.24 & 0.76 & 22.19 \\
PKS0312-77~~\#045 & 2.00 & 0.61 & 21.95 \\
PKS0312-77~~\#031 & 1.63 & 0.56 & 21.91 \\
PKS0312-77~~\#036 & 1.84 & 0.63 & 22.01 \\
PKS0537-28~~\#111 & 1.97 & 0.47 & 22.00 \\
PKS0537-28~~\#054 & 1.47 & 0.38 & 22.08 \\
PKS0537-28~~\#091 & 1.57 & 0.57 & 22.10 \\
PKS0537-28~~\#037 & 1.28 & 0.53 & 22.13 \\
\noalign{\smallskip}\hline
\end{tabular}
\end{minipage}
\end{table}
%
 \subsection{Optical Imaging and Source Cross-matching}
 \par\noindent
 The optical imaging ($R$--$Bessel$ filter) complementary to our deep
 near-infrared observations was obtained during two runs using EFOSC2
 at the ESO 3.6m telescope, in December 2000 and August 2001.
 Typical exposure times were about 10~min per field.
 The images were reduced using standard IRAF routines.
 For each observing run, we obtained the proper bias by
 combining several bias frames with a median filter. The images
 were then corrected for pixel to pixel response variations
 using the median of several flat-field frames obtained during 
 the night. Cosmic rays and bad pixel corrections were applied
 to each frame. 
 The photometric calibration was performed for each night
 using the zero-point derived from the measured instrumental magnitudes
 of standard stars and assuming the average extinction reported in the
 Observatory web pages\footnote{http://www.eso.org/observing/support.html.}.
 Object detection was performed using \sex.  
 Since the images were obtained under different seeing conditions
 (from 1\farcs1 to 2\farcs1),
 we used different threshold values 
 from image to image so that the faintest detectable sources in all frames
 have a minimum signal--to--noise ratio of 3 over the seeing disk.
 Using these detection parameters, 
 none of the hard X-ray selected sources discussed here was detected
 in the $R$~band, and the limiting magnitude was estimated for each of
 them as the 3$\sigma$ value over 2.5 times the seeing area.
 These limiting magnitudes (24.5 $< R_{lim} <$ 25.2) were adopted 
 to compute the lower limits of X/O reported in Table~\ref{tabsam}.
 All of the magnitudes were corrected for Galactic extinction using
 the values from the NED Galactic extinction
 calculator\footnote{http://nedwww.ipac.caltech.edu/forms/calculator.html.}
 based upon the maps of Schlegel, Finkbeiner \& Davis (\cite{galext}),
 although the extinction is generally small in our fields.


 In order to cross match the near-infrared and optical catalogs, we elected
 as reference the \ks \ filter. First, we positionally registered the $R$
 images to the \ks \ mosaics, using $\approx 10-15$ bright compact objects
 (mainly stars) for each frame. After having brought the two independently
 produced catalogs to a common astrometric reference, for each entry in the
 \ks~catalog, we searched for objects closer than 2\arcsec \ in the
 $R$~catalog. In 80\% of the cases we found an $R$-counterpart for entries
 brighter than \ks = 20.5 in the near-infrared catalog:
 the mean offset is consistent with zero, while the typical r.m.s. of the 
 position differences in both right ascension and declination are of the order
 of 0.4\arcsec. 
 At this stage we can confidently measure the colors of the galaxies in fixed
 apertures of 3\arcsec-diameter. We first degraded the seeing of the \ks \
 images to that of the $R$~band one, then we measured
 aperture magnitudes on both the $R$ and \ks \ images (using the IRAF
 task {\tt PHOT}) at the position given by the original detections on $ISAAC$
 data. 
 With this method we increased the percentage of \ks \ entries with a measured
 \rk \ color with respect to simply cross-correlating the two independently
 produced catalogs. Most important, we can measure,
 although with large errors, the $R$ magnitude for a fraction of the
 \ks-detected counterparts of our X-rays targets.
 These sources were originally absent from the optical catalog, being
 just below the detection threshold limits.
%
%
\section{Properties of the IR Counterparts}
\begin{table*}
       \caption{Near Infrared Properties}
       \label{tabpropir}
       \vspace{0.2cm}
\begin{minipage}{0.99\textwidth}
\begin{tabular}{lllclllr}
\hline\hline\noalign{\smallskip}
 SOURCE ID. & RA(\ks) & DEC(\ks) & 
 $\Delta$(\ks-X)\footnote{Positional offset between the near-infrared counterpart and X-ray sources.} & 
 \ks\footnote{``total'' magnitude: MAG\_BEST from {\it SExtractor}.} & 
 \rk\footnote{Fixed 3\arcsec-diameter aperture color.} &
 $P$(field)\footnote{Probability of the galaxy being located near the X-ray source by chance.} &
 $P$(EROs)\footnote{Probability of being an Extremely Red Object located near the X-ray source by chance.}\\
  & (J2000) & (J2000) & (\arcsec) &   &   & \\
\noalign{\smallskip}\hline\noalign{\smallskip}
Abell2690~~~\#075 & 23 59 56.41 & -25 10 17.6 & 3.5 & 18.32 & $6.3\pm0.7$ & $2.2\times10^{-2}$& $1.1\times10^{-4}$\\ 
Abell2690~~~\#029 & 00 01 11.50 & -25 12 06.5 & 3.8 & 17.68 & $7.4\pm1.0$ & $1.6\times10^{-2}$& $1.0\times10^{-4}$\\ 
                  & 00 01 11.49 & -25 12 08.7 & 5.9 & 18.48 & $>5.9     $ & $7.3\times10^{-2}$& $3.0\times10^{-4}$\\ 
BPM16274~~\#069   & 00 50 30.86 & -52 00 47.6 & 2.2 & 17.83 & $6.6\pm0.8$ & $4.2\times10^{-3}$& $3.0\times10^{-5}$\\ 
BPM16274~~\#181   & 00 50 31.51 & -52 06 33.9 & 3.0 & 18.69 & $6.2\pm0.4$ & $2.6\times10^{-2}$& $3.6\times10^{-4}$\\ 
                  & 00 50 31.35 & -52 06 34.2 & 3.8 & 19.43 & $>5.3$      & $7.2\times10^{-2}$& $1.4\times10^{-2}$\\ 
PKS0312-77~~\#045 & 03 10 18.93 & -76 59 58.4 & 0.5 & 18.70 & $5.7\pm0.7$ & $6.4\times10^{-4}$& $4.6\times10^{-5}$\\ 
                  & 03 10 17.93 & -76 59 57.9 & 3.6 & 19.45 & $>5.5$      & $6.2\times10^{-2}$& $1.2\times10^{-2}$\\ 
PKS0312-77~~\#031\footnote{source in the FOV of {\it Chandra} observations.}  
& 03 11 13.65 & -76 53 59.6 & 1.0 & 18.31 & $5.3\pm0.5$ & $1.7\times10^{-3}$& $9.6\times10^{-5}$\\ 
PKS0312-77~~\#036$^f$
& 03 13 42.90 & -76 54 23.9 & 2.9 & 19.11 & $5.5\pm0.5$ & $3.2\times10^{-2}$& $2.9\times10^{-3}$\\ 
\hfill\footnote{spectroscopically confirmed star, but 4\farcs8 from the {\it Chandra} X-ray source.} 
		  & 03 13 43.76 & -76 54 19.6 & 3.5 & 18.67 & $2.3\pm0.2$ & $3.1\times10^{-2}$& \\
                  & 03 13 42.68 & -76 54 26.7 & 5.5 & 18.71 & $5.1\pm0.5$ & $7.4\times10^{-2}$& $5.5\times10^{-3}$\\ 
PKS0537-28~~\#111$^f$ 
& 05 39 11.60 & -28 37 15.1 & 3.2 & 17.66 & $6.8\pm0.7$ & $1.1\times10^{-2}$& $9.0\times10^{-5}$\\ 
                  & 05 39 11.35 & -28 37 16.7 & 2.2 & 20.11 & $4.6\pm1.0$ & $3.5\times10^{-2}$& \\
PKS0537-28~~\#054$^f$ 
& 05 39 45.22 & -28 49 09.8 & 1.6 & 18.91 & $>6.2$      & $8.0\times10^{-3}$& $1.2\times10^{-4}$\\ 
PKS0537-28~~\#091$^f$ 
& 05 40 21.13 & -28 50 37.6 & 1.0 & 18.99 & $5.1\pm0.5$ & $3.3\times10^{-3}$& $4.3\times10^{-4}$\\ 
\noalign{\smallskip}\hline
\end{tabular}
\end{minipage}
\end{table*}
%
 \subsection{Candidates Identification and Reliability}
 \par\noindent
 On the basis of the optical identification of the whole \hel \ survey
 (Fiore et~al. \cite{H2X1dF}; Brusa et~al. \cite{p0312}), we know that all
 of the counterparts of our X-ray sources are within 6\arcsec \ from the
 \xmm \ position (actually within 3\arcsec \ for $\sim$80\% of the cases).
 We therefore searched in the \ks~catalog for objects closer than
 6\arcsec \ to the X-ray position: \emph{in 10 out of 11 cases we found a
 relatively bright (\ks $\leq$ 19) near-infrared source well within the
 error circle} (see Fig.\ref{fig_fc}). 
 The bright near-infrared counterparts have optical to
 near-infrared colors considerably redder than the field population;
 indeed, all of them are classified as Extremely Red Objects
 (EROs, with \rk $>5$).
 The only exception is PKS0537-28~\#037, where no object is detected
 within the error box, but several relatively bright galaxies,
 with similar \rk \ colors, are present at distances less than 10\arcsec \
 from the nominal X-ray position: in this case the X-ray flux 
 is probably due to thermal emission from the hot intragroup gas. 
 Nevertheless, we cannot exclude the possibility that the
 true counterpart of this X-ray source might be either heavily reddened 
 (e.g. similar to some AGN--hosting SCUBA galaxies, Ivison et~al.
 \cite{ivison}) or at very high redshift (Koekemoer et~al. \cite{koeke03}). 

 In five cases more than one object is detected in the \ks~band
 within 6\arcsec \ from the X-ray centroid.
 Table~\ref{tabpropir} gives all the relevant photometric information
 about the detected counterparts.
 We decided to adopt a statistical approach to securely identify
 the near-infrared object responsible for the X-ray emission.
 For each counterpart detected within 6\arcsec \
 from the X-ray position we computed the probability
 to find by chance an object within this radius:
 
   \begin{equation}
    P = 1.0-\exp{(-\pi \Delta^2 n(<K_s))},
    \end{equation}
 
 where $\Delta$ is the distance between the X-ray and
 near-infrared sources, and $n(<K_s)$ the integral surface
 density of galaxies brighter than \ks. 
 If more than one counterpart is present, the most likely 
 candidate is that with the lowest value of $P$ 
 (see Downes et~al. \cite{downes} for more details).
 Two sets of probabilities are reported in Table~\ref{tabpropir},
 which were computed assuming the surface density of field galaxies 
 and EROs, respectively. The surface density of field galaxies in
 the \ks \ band was determined from our data 
 and without taking into account any color information, 
 whereas for the Extremely Red Objects (with \rk $>5$)
 we used the results
 of Daddi et~al. (\cite{daddifield}) and Roche et~al. (\cite{roche02}). 
 In all but one of the cases, where more than one counterpart is present 
 within the X-ray error circle, the brightest
 \ks \ counterpart is also the nearest. 

 Five of our targets in the {\tt PKS0312-77} and {\tt PKS0537-28} fields
 were in the field of view of {\it Chandra} observations.
 All five sources are clearly detected by a 
 standard detection algorithm (we refer to Brusa et~al. \cite{p0312} for 
 details on the {\it Chandra} data reduction).
 The {\it Chandra} positions are shown in Fig.\ref{fig_fc} with a cross.
 In all cases, the better positional
 accuracy of {\it Chandra} unambiguously points to the reddest, and
 near-infrared brighter object, with an offset always smaller than 1\farcs5.
 Thanks to the excellent position accuracy of {\it Chandra} observations 
 it was possible to secure the correct identification of the source
 {\tt PKS0312-77~\#36}, where three objects with similar chance
 probabilities were found in the \xmm \ error box. 

 The positional coincidence with the most probable candidate when a
 {\it Chandra} observation was available, allowed us to
 unambiguously elect the most probable candidate as responsible of the
 hard X-ray emission for all of the sources in the sample. 
 
 The presence in four cases of another object with ERO colors 
 (and for one source, {\tt BPM16274~\#181}, the presence of five
 EROs within 5\arcsec \ from the brighter \ks \ counterpart,
 indicating a possible high-$z$ cluster),
 is a natural consequence of the already known clustering properties
 of these objects (Daddi et~al. \cite{K20_2}). We refer to these objects
 as ``secondary'' counterparts.
 
 Summarizing, we have positively identified 10 hard X-ray, high X/O sources
 with relatively bright \ks \ counterparts; all of them are Extremely Red
 Objects and most of them (6 out of 10) belong to the extreme tail of the
 ERO population (\rk $>6$). 
%
   \begin{figure*}
   \centering
   \includegraphics[width=0.79\textwidth]{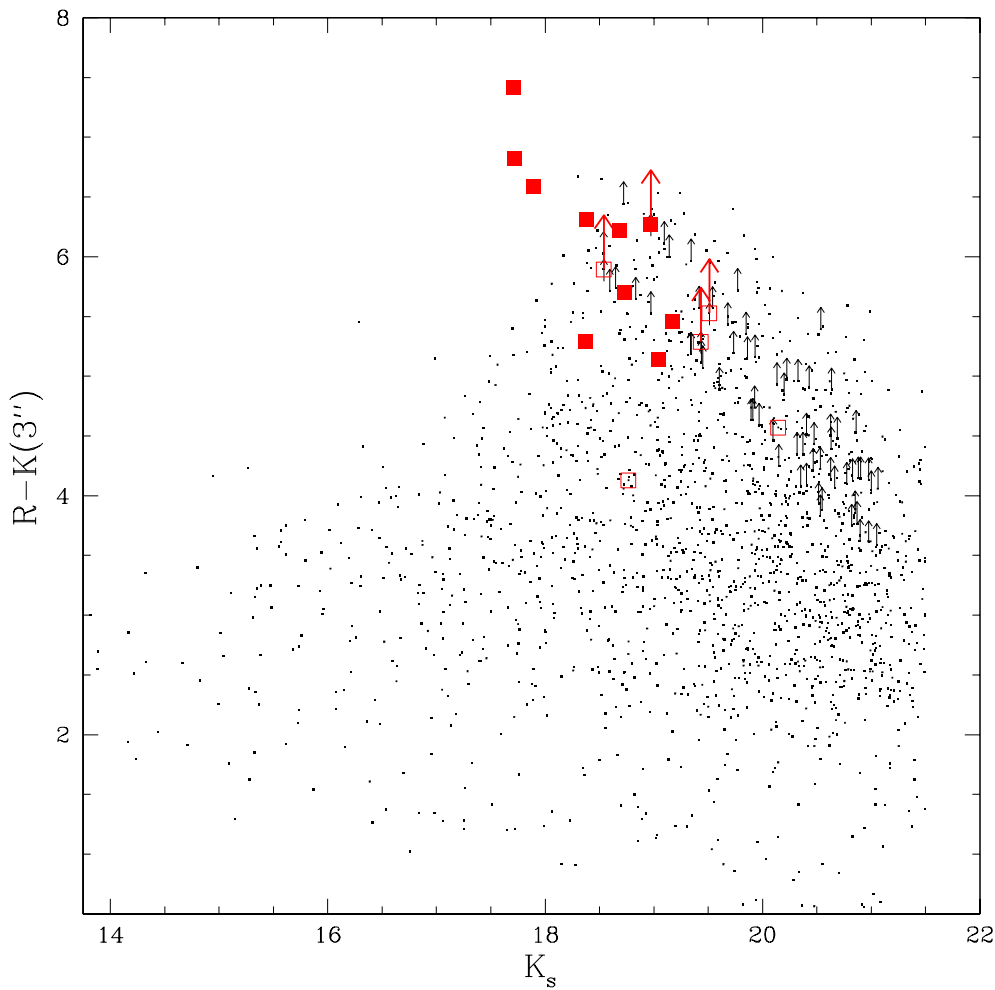}
   \caption{Fixed-aperture $R$-$K_s$ color versus 
   ``total'' \ks \ magnitude for all the {\it ISAAC\/} detections (small dots)
   with \ks$<21.5$; the large filled squares indicate the ten counterparts
   of the selected hard X-ray sources with high X/O. The large empty
   squares are the ``secondary'' counterparts.
   }
    \label{fig_K_RK}
    \end{figure*}
%
%
\subsection{Comparison with Field Population}
 \par\noindent
 The \rk \ color distribution of all of the
 {\it ISAAC\/} detections with \ks$<21.5$ is shown in 
 Fig.\ref{fig_K_RK}. The exceptional nature of the 
 high X/O counterparts is quite obvious:
 the three brightest counterparts, with \ks$<18$ \emph{and}
 \rk $>6$, are unique with respect to the galaxy field population,
 as no other object shares the same photometric properties.
 The six \rk $>6$ counterparts, with \ks$<19$, are also very rare objects:
 in this region of the color-magnitude
 plane we find 20 galaxies in total, representing a mere 3\% of the
 whole galaxy population\footnote{This fraction is higher than the value
 (1.7\%) found by Daddi et~al. (\cite{daddifield}) in their larger survey,
 but we must take into account the fact that our fields can not be
 considered random fields with respect to the ERO surface density.}.
 All of the 10 most likely counterparts satisfy the ERO color criterion
 \rk $>5$ (and 9 out of 10 also the more stringent limit
 \rk $\ge 5.3$) and are brighter than \ks=19.2. Given the ERO surface density
 expected at these magnitude levels ($\sim$1 object per square arcmin,
 see Daddi et~al. \cite{daddifield}; Roche et~al. \cite{roche02}), it is
 remarkable that we found them nearly ubiquitously in our small error
 boxes.

 The ERO population selected in the optical/near-infrared
 surveys is known to consist of old, passively evolving ellipticals and
 star-forming galaxies strongly reddened by dust extinction at high
 redshift ($z\,\gsimeq\,1$). The two components seem to be about
 equally represented
 (see Mannucci et~al. \cite{mannucci} for a tentative photometric separation
 or Cimatti et~al. \cite{K20_1} for the first spectroscopic classification
 of an ERO sample). In addition, a few examples of EROs hosting AGN 
 have been discovered so far (Afonso et~al. \cite{afonso};
 Pierre et~al. \cite{pierre}).
 Finally, red quasars with  very large optical-to-near infrared colors
 (up to \rk $\sim$8) have been discovered in radio surveys
 (Webster et~al. \cite{webster}; Lawrence et~al. \cite{lawrence};
 Gregg et~al. \cite{gregg}).
 
 In order to discriminate among the above mentioned possibilities,
 and to constrain the source redshift, the exceptional quality of 
 the {\it ISAAC\/} images has been fully exploited by combining the
 photometric information with an accurate morphological analysis.

%
\subsection{Near-Infrared Morphologies}
   \begin{figure*}
   \centering
   \includegraphics[width=0.79\textwidth]{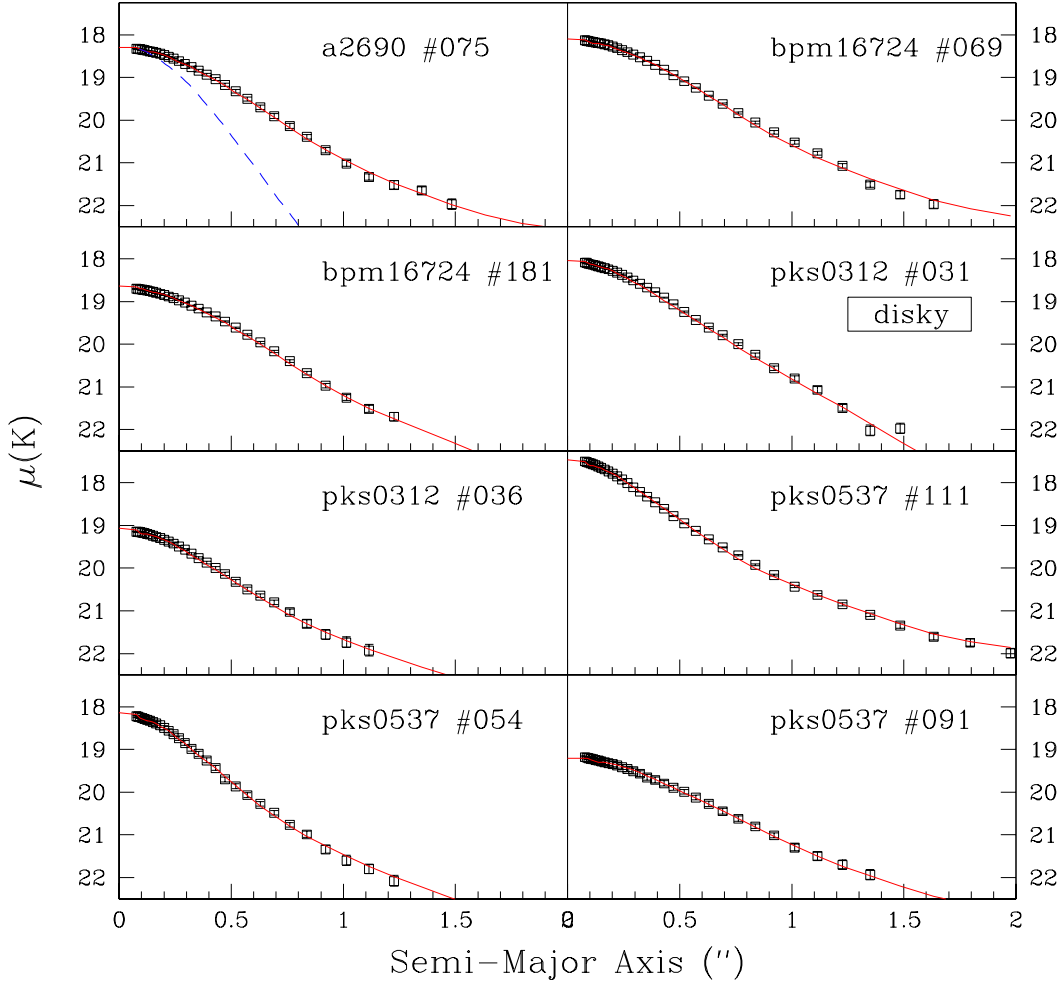}
   \caption{Surface brightness profiles along the major axis (boxes) and
   best-fitting models (lines) for all of the extended sources. In the upper
   left box we also show with a dashed line a representative PSF.}
    \label{fig_prof}
    \end{figure*}
%
\par\noindent
 The morphologies of the near-infrared counterparts were 
 investigated  using the IRAF task {\tt isophote}, which fits each galaxy
 with elliptical isophotes, providing a radial intensity profile along the
 semi--major axis.  With the exception of the 
 identifications of {\tt Abell2690~\#029} and {\tt PKS0312$-$77~\#045},
 all of the other sources are clearly extended. 

 From the analysis of the radial profiles the half-light radii
 ($r_{hl}$) were computed. In a few cases, a nearby
 object can affect the profile fitting: in these cases we have masked out
 the neighbors before carrying out the analysis. 
 A least-square fitting, excluding the inner region affected by the seeing,
 was performed in order to distinguish between exponential and bulge
 ($r^{1/4}$) profiles. This straightforward approach, which simply uses
 $r_{hl}$, a \emph{measured} model-independent parameter, 
 allowed us to assign a morphological type to all but one
 of the galaxies. Only for the counterpart of {\tt BPM16274~\#069} both
 the exponential and $r^{1/4}$--law profiles give equally 
 acceptable fits. The good quality of the fits, which use only ``canonical''
 galaxy profiles, is a first clear indication that any contribution from
 unresolved sources, which would be expected to trace the 
 X-ray emitting AGN, is probably negligible in the extended sources. 
 The results are reported in Table~\ref{tabmorfir}.

\begin{table*}
       \caption{Near Infrared Morphological Parameters}
       \label{tabmorfir}
       \vspace{0.2cm}
\begin{minipage}{0.99\textwidth}
\begin{tabular}{lllclclc}
\hline\hline\noalign{\smallskip}
 SOURCE ID. &
 \ks\footnote{``total'' magnitude: MAG\_BEST from {\it SExtractor}.} & 
 \rk\footnote{Fixed 3\arcsec-diameter aperture color.} &
 morph\footnote{From least-square fit: P=point-like -- E=Bulge profile --
 D=Disky profile.} &
 $r_{hl}$ & morph\footnote{From simulations.} & $r_{eff}$ &
 gal/nuc\footnote{Ratio between the host galaxy and the unresolved
 nuclear component fluxes.}\\
  & & & & & & &\\
\noalign{\smallskip}\hline\noalign{\smallskip}
Abell2690~~~\#075  & 18.32 & $6.3\pm0.7$ & (E) & 0\farcs56 & E & 0\farcs56 & $>40$ \\
Abell2690~~~\#029  & 17.68 & $7.4\pm1.0$ &  P  \\
BPM16274~~\#069    & 17.83 & $6.6\pm0.8$ & E/D & 0\farcs60 & E & 0\farcs48 & $>50$ \\
BPM16274~~\#181    & 18.69 & $6.2\pm0.4$ & (E) & 0\farcs53 & E & 0\farcs37 &~~$6.3$ \\
PKS0312-77~~\#045  & 18.70 & $5.7\pm0.7$ &  P  \\
PKS0312-77~~\#031  & 18.31 & $5.3\pm0.5$ &  D  & 0\farcs55 & D & 0\farcs47 &~~$1.9$ \\
PKS0312-77~~\#036  & 19.11 & $5.5\pm0.5$ & (E) & 0\farcs55 & E & 0\farcs57 & $>18$ \\
PKS0537-28~~\#111  & 17.66 & $6.8\pm0.7$ &  E  & 0\farcs61 & E & 0\farcs70 &~~$33.$ \\
PKS0537-28~~\#054  & 18.91 & $>6.2$	&  E  & 0\farcs43 & E & 0\farcs48 &~~$7.9$ \\
PKS0537-28~~\#091  & 18.99 & $5.1\pm0.5$ &  E  & 0\farcs66 & E & 0\farcs80 & $>25$ \\

\noalign{\smallskip}\hline
\end{tabular}
\end{minipage}
\end{table*} 

 The half--light radius determination may be biased by several effects. 
 The most important are the seeing blurring and the effect
 of the surface brightness threshold imposed by the noise 
 ($\sigma_{\mu(k_s)}\approx$ 22~mag/arcsec$^2$; see Table~\ref{tabobsir})
 in computing the total magnitude.
 In order to correct for these effects, and hence to obtain a more reliable
 measure of the \emph{intrinsic} effective radius of our galaxies,
 we performed accurate simulations using the {\tt IRAF} task {\tt mkobject}.
 A grid of models including a point-like
 source and two sets of bulge and disc (exponential) profiles with
 a wide range of effective radii and flux ratios between the point-like
 and the extended components were computed. 
 The simulations were also convolved with the seeing, where the 
  PSF has been modeled with a Moffat function with parameters 
 determined using about $5-10$ stars per field. 
 As for the real objects, we use 
 {\tt isophote.ellipse} to extract a radial intensity profile for
 each model. The model giving the smallest residuals compared to
 the observed profile (after checking by eye the two-dimensional
 ``object$-$model'' image) is then used to estimate the
 morphological type, the ``seeing-corrected'' effective radius, and
 the possible contribution from an unresolved nuclear component.
 In the last three columns of Table~\ref{tabmorfir}, the best--fit results
 from the  simulations are given, showing an excellent agreement with those
 of the direct profile fitting: all of the morphological types have been
 confirmed, and the object with uncertain classification 
 ({\tt BPM16274~\#069}) is now consistent 
 with an elliptical profile. Moreover, in all but
 one case the possible contribution from a central, unresolved source
 is almost negligible (\lsimeq 15\%), the only exception being the source 
 {\tt PKS0312$-$77~\#031}, a disky galaxy for which the simulations admit the
 presence of a point-like nucleus, about half as bright as the host galaxy.
 The best-fitting models superimposed on the near-infrared surface brightness
 profiles are presented in Fig.\ref{fig_prof}.
 The results of the simulations allow us to estimate how the observational
 biases affect the half--light radius determination:
 for the exponential profiles\footnote{The simulations included also 
 the ``secondary'' counterparts of {\tt Abell2690~\#029} and
 {\tt PKS0312$-$77 \#036}, and both 
 are disky galaxies.}, the intrinsic effective radius 
 derived from the simulations is always smaller than that measured directly
 on the images, since these profiles are more
 prone to seeing blurring. The same behavior is present also in the
 elliptical hosts observed under the worst seeing conditions (in
 both the BPM16274 fields), whereas the intrinsic effective radius 
 is larger than the measured half-light radius for those objects
 observed in optimal seeing conditions.
 In the latter case, the dominant observational effect is the light loss
 in the outermost regions of the galaxy profile, which 
 leads to an underestimate of the total luminosity, thus affecting the
 half-light radii determination. This effect is due to the surface
 brightness threshold imposed by the noise, which is known to
 be more important for the slowly decreasing elliptical profiles
 (Cimatti et~al. \cite{K20_3}, Angeretti, Pozzetti \& Zamorani \cite{rosso}).
%
%
\section{Discussion}
%
   \begin{figure*}
   \centering
   \includegraphics[width=0.77\textwidth]{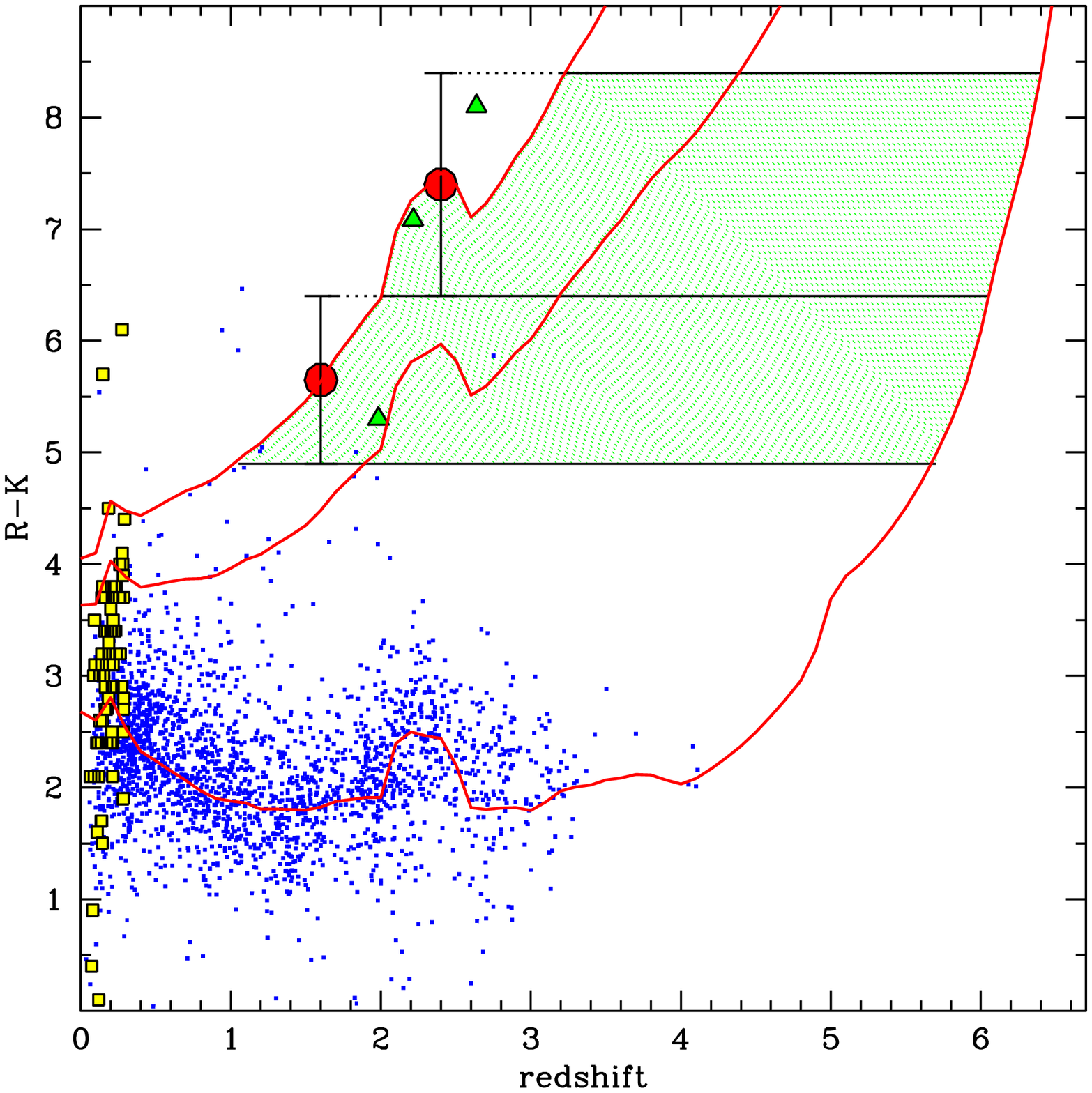}
   \caption{Redshift vs. optical-to-near infrared color for 
   pointlike objects. The large filled dots represent our point-like sources.
   The small squares are the optically selected quasars from 
   Barkhouse \& Hall (\cite{bark}), whereas the larger squares indicate 
   the 2MASS quasar sample (Hutchings et~al. \cite{2MASSQSO}). 
   The three solid triangles are the high-$z$ reddest QSO from
   Gregg et~al. (\cite{gregg}). 
   The curves represent the color--redshift relations for the
   adopted quasar template with three different extinction
   values ($E(B-V)=0.0,0.4,0.7$, from bottom to top).
   The redshift-permitted region of \hel \ sources is shaded. 
   }
    \label{fig_z_RK_point}
    \end{figure*}
%
%
\subsection{Redshift constraints}
%
\par\noindent
Because of their high X/O ratios and large \rk \ colors,
we argue that most of the sources are likely to be obscured AGN at high
redshift. However, the lack of a multi-band coverage does not allow
us to estimate a precise photometric redshift. Broad-band
photometry in at least 4-6 filters (from $U$ to $K$) would be needed
for a reliable redshift estimate for these objects (Bolzonella et~al.
\cite{micol})\footnote{We note, however, that due to the extreme faintness
of these objects in the UV/optical region, a multi-color observation would
be extremely time consuming.}. Nevertheless, we have estimated a ``minimum''
photometric redshift (\zmin) for all of the sources 
using the color-redshift diagram \hbox{(\rk \ vs. $z$)} 
and different templates according to their morphology.

In particular for the point-like sources,
since the emission from the AGN contributes significantly
in the optical/near-IR range, we have used as a QSO template 
the composite spectrum from the Large Bright Quasar Survey
(LBQS; Francis et~al. \cite{francis}), extended to the near-IR 
using the mean radio-quiet quasar energy distribution by
Elvis (\cite{elvised}; see also Maiolino et~al. \cite{maio00}).
Conversely, for extended objects (both bulge and disky morphology,
respectively E and D in Table~\ref{tabmorfir}) the minimum photometric
redshifts are based on the assumption that their 
optical/near-infrared spectral energy distribution (SED) is due to
stellar processes (this assumption is strengthened by the results on
the low contribution of point-like nuclei in these counterparts).
The adopted templates consist
of a set of synthetic spectra from GISSEL 2000 (Bruzual \& Charlot
\cite{bruzual}), with different star formation histories
and spanning a wide range of ages; 
the basic set of templates includes only solar metallicity
and Salpeter's Initial Mass Function (IMF, Salpeter \cite{salpeter}).

In Figure~\ref{fig_z_RK_point} we show the color-redshift diagram
for our two point-like sources (large filled dots) along with data taken
from the literature. We have included the optically selected quasar sample
with near-IR counterparts in the 2MASS (Barkhouse et~al. \cite{bark};
small squares), which nicely fits the extended LBQS template.
The effect of the intergalactic medium (IGM) attenuation (Madau et~al.
\cite{madau}) is clearly visible at high redshift. Also shown is the
2MASS--selected sample of red quasars from Hutchings et~al. (\cite{2MASSQSO})
at low redshift (large squares), and the high-$z$ reddest QSOs from
Gregg et~al. (\cite{gregg}; solid triangles).
In order to reproduce the color of red quasars, we have included the effect
of internal dust attenuation on the Francis extended template, 
using a dust--screen model and the SMC extinction law (Pei \cite{pei}).
The attenuated templates reproduce quite well most of the low-$z$
red quasar sample, as well as the high-$z$ reddest QSOs, 
with E(B-V) ranging from $0$ to $0.7$, corresponding to $A_V<2.0$ 
(see also Gregg et~al. \cite{gregg}).
First, we note that most (10 over 13) of observed QSOs with 
\rk $>5$ lie at $z>1$, 
the only exceptions being one optically selected quasar
(Barkhouse et~al. \cite{bark}) and two 2MASS selected QSOs at $z<0.3$
(Hutchings et~al. \cite{2MASSQSO}). 
However, the X-ray properties of the low-$z$ 2MASS QSOs 
are completely different from those of the sources in our sample, 
having on average very low X/O ratios (Wilkes et~al. \cite{wilkes}).
Therefore, from the analysis of literature data we could
confidently infer that our extremely red point-like sources lie at $z>1$. 
In addition, we use the upper envelope of the observed
color--redshift diagram for spectroscopically identified QSO, 
reproduced by an extinction value $E(B-V)\simeq0.7$ (in agreement
with the maximum confirmed AGN extinction to date; Gregg et~al. \cite{gregg}),
to estimate the minimum redshift for a given \rk \ color.
We can thus assert, with good confidence, that the two point-like sources
in our sample are at $z>1$ and derive their \zmin \
(see Table~\ref{tabzmin}).
The observed colors correspond to $z_{\rm min}=1.65^{+0.35}_{-0.65}$ for 
{\tt PKS0312$-$77~\#045} and $z_{\rm min}=2.40^{+0.90}_{-0.40}$ for
{\tt Abell2690~\#029A}.
The rest--frame de--reddened luminosities in the B\footnote{the effect of
the dust in the B-band is about $2.6$ magnitude for $E(B-V)=0.7$.}
and \ks \ bands are consistent with the optical and near-IR luminosity
distributions in the Barkhouse et~al. (\cite{bark}) sample 
($\langle M_B \rangle=-26.3$, $\sigma(M_B)=2.2$ 
and $\langle M_K \rangle=-29.4$ , $\sigma(M_K)=1.9$). 
Finally, from the unattenuated QSO template, we can also place a limit
to the maximum redshift of point-like sources
(see Fig.\ref{fig_z_RK_point}).
Indeed, assuming that their extremely red colors are due to the high-$z$
IGM absorption, their maximum redshift is around $z=6-6.5$.
In this case the luminosities derived from the near-IR
are typically 1$-$1.5 magnitudes brighter than in \zmin \ case.
   \begin{figure*}
   \centering
   \includegraphics[width=0.77\textwidth]{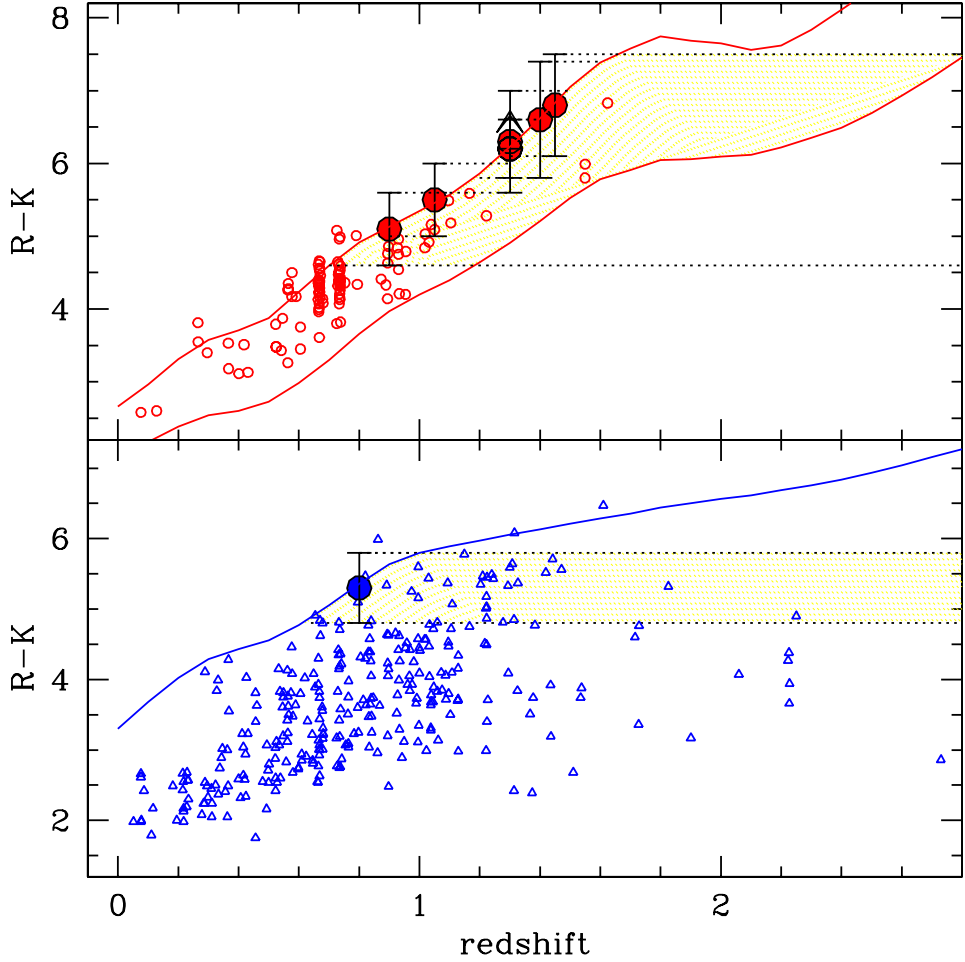}
   \caption{Redshift vs. optical-to-near infrared color for 
   extended objects. The large filled symbols represent the \hel \ sources.
   The smaller empty symbols are sources from the K20 spectroscopic
   survey (Cimatti et~al. \cite{K20_3}). In the upper panel we compare our
   bulge profile sources with the K20 early-type galaxies. In the lower panel
   we compare the disky sources with a sample of emission line galaxies.
   In both panels also the ``maximum color -- minimum redshift'' track
   (see text) is shown.
   The redshift-permitted region of \hel \ sources is shaded. 
   }
    \label{fig_z_RK_extd}
    \end{figure*}
%

With a similar procedure we have estimated a \zmin \ for
sources with a bulge or disky profile.
For the bulge-profile sources we have used for comparison the
sub-sample of early-type galaxies in the K20 spectroscopic survey
(Cimatti et~al. \cite{K20_3}). In Figure~\ref{fig_z_RK_extd} we show
the model which best represents the upper envelope 
of the K20 data of early-type galaxies, which is computed assuming 
a simple stellar population forming at high redshift ($z_{\rm form}=20$).
It corresponds to a ``maximum age'' model and provides the reddest colors
at each redshift. A lower $z_{\rm form}$ and/or a more
gradual star formation history would result in bluer colors and in 
a corresponding higher \zmin.
For the disky sources we have used for comparison the sub-sample of 
emission-line galaxies of the K20 survey. To reproduce the colors
of the most extreme emission-line dusty objects we have chosen
a model with constant Star-Formation Rate (SFR), and a SMC
extinction law with $E(B-V)=0.6$
(cf. Cimatti et~al. \cite{K20_3} and Pozzetti et~al. \cite{K20_5}).
We have then assigned a \zmin \ to each of our sources according
to their \rk \ color and morphological type.

We found that the minimum redshifts for bulge and disky profile sources
are in the range of $0.80 \le z \le 1.45$.
In Table~\ref{tabzmin} we list the \zmin \ and their color-related
uncertainties. Also reported are the intrinsic rest--frame luminosities
in the $B$ and \ks \ bands.
We note that most of the extended sources are more than 1 magnitude 
brighter than typical local $L^*_K$ galaxies ($M^*_{K_s}=-24.21$  at $z=0$
from Cole et~al. (\cite{cole})) and similar or even brighter than
$L^*_K$~galaxies at $z=1-1.5$ (Pozzetti et~al. \cite{K20_5}). 
Taking into account that the $M_{stars}/L_K$ ratio for an old stellar
population could vary from 0.3 (for an age of 1 Gyr, assuming a Salpeter IMF)
to about 1 (for 6~Gyrs, the age of the universe at~$z=0.9$), 
we estimate the galaxy stellar Masses of our hosts: they are all
above $3\times 10^{10}$ M$_\odot$ (and 4 of them above $10^{11}$ M$_\odot$),
corresponding to the stellar-Mass of an $L^*$ galaxy 
in the local Universe (Cole et~al. \cite{cole}).
The elliptical hosts of hard X-ray sources with extreme X/O are therefore
among the most massive spheroids at these redshifts, possibly addressing the
issue of elliptical galaxy formation and the expected co-evolution
with the black-hole accretion (Granato et~al. \cite{granato}).

The procedure above described leaves the maximum redshifts basically
unconstrained. However, we can infer $z_{\rm max}$
for bulge sources assuming a lower bound to the observed K20 data,
represented by an old stellar population with an age of $1$ Gyr, consistent
also with their relaxed morphological properties. The expected range in
redshift ($z_{min} - z_{max}$) is shown by the shaded area in
Fig.\ref{fig_z_RK_extd}.

\begin{table*}
	\caption{Minimum redshifts and intrinsic rest-frame properties}
	\label{tabzmin}
	\vspace{0.2cm}
\begin{minipage}{0.99\textwidth}
\begin{tabular}{lccccccc}
\hline\hline\noalign{\smallskip}
 ID & morph & $z_{min}$ & $M_K$ & $M_B$ & 
 $r_e$ & N$_H$ & L$_{2-10 {\rm keV, unabs}}$ \\
  & & & & & (kpc) & (10$^{22}$ cm$^{-2}$) &  (10$^{44}$ erg s$^{-1}$) \\
\noalign{\smallskip}\hline\noalign{\smallskip}
%
Abell2690~~\#075   & E & $1.30^{+0.20}_{-0.20}$	& $-25.83$ & $-21.77\;$ & $4.7$  & 15.3$^{+23.2}_{-\;9.1}$  &  3.5 \\
Abell2690~~\#029  & P & $2.40^{+0.90}_{-0.40}$	& $-29.74$ & $-26.31\footnote{de--reddened by 2.6 mag assuming $E(B-V)=0.7$.}$ 
                                                                        & $-$    &  2.8$ ^{+\;3.5}_{-\;2.1}$  &  11.2\\
BPM16274~~\#069    & E & $1.40^{+0.20}_{-0.20}$ & $-26.53$ & $-22.49\;$ & $4.1$  &  2.6$^{+\;1.6}_{-\;1.1}$ &  3.6 \\
BPM16274~~\#181    & E & $1.30^{+0.10}_{-0.10}$	& $-25.45$ & $-21.40\;$ & $3.1$  &  1.1$^{+\;3.2}_{-\;1.0}$ &  1.2 \\
PKS0312-77~~\#045  & P & $1.65^{+0.35}_{-0.65}$ & $-28.02$ & $-24.58^a$ & $-$    &  6.6$^{+\;7.4}_{-\;3.8}$ &  4.9 \\
PKS0312-77~~\#031  & D & $0.80^{+0.20}_{-0.20}$ & $-24.77$ & $-22.10\footnote{
de--reddened by 2.3 mag assuming $E(B-V)=0.6$}$ & $3.5$  & 1.2$^{+\;1.2}_{-\;0.8}$     &  0.5\\
PKS0312-77~~\#036  & E & $1.05^{+0.20}_{-0.20}$	& $-24.44$ & $-20.33\;$ & $4.6$  &  1.2$^{+\;1.5}_{-\;1.0}$ &  1.0 \\
PKS0537-28~~\#111  & E & $1.45^{+0.20}_{-0.20}$	& $-26.80$ & $-22.77\;$ & $5.9$  & 12.0$^{+16.6}_{-\;7.2}$   &  2.8 \\
PKS0537-28~~\#054  & E & $>1.30$ 		& $-25.24$ & $-21.18\;$ & $4.0$  &  1.6$^{+\;2.4}_{-\;1.2}$    &  $>1.6$ \\
PKS0537-28~~\#091  & E & $0.90^{+0.20}_{-0.20}$	& $-24.17$ & $-20.01\;$ & $6.2$  & 15.7$^{+49.9}_{-11.2}$    &  2.0 \\
\noalign{\smallskip}\hline
\end{tabular}
\end{minipage}
\end{table*}

%
\subsection{Angular Sizes of the EROs}
%
   \begin{figure}
   \centering
   \includegraphics[width=0.49\textwidth]{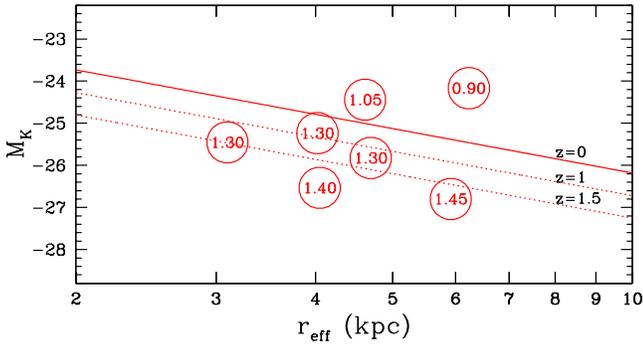}
   \caption{Effective radius vs. luminosity relation for extended objects
   with a bulge profile. Objects are identified by their $z_{min}$. 
   The solid line shows the local relation from Pahre et~al. (\cite{pahre}), 
   while the dotted lines show its evolution at $z=1$ and $z=1.5$, assuming
   the luminosity evolution derived by Pozzetti et~al. \cite{K20_5}.
   }
    \label{fig_re_LK}
    \end{figure}
%

\par\noindent

In section 4.3 we have shown that 
our accurate morphological profile analysis has allowed us to measure
the effective angular radius for the extended sources.
First of all, we note that their angular sizes are in general quite small,
ranging from $0.5\arcsec$ to $0.8\arcsec$, and therefore
consistent with being high-$z$ objects.
For a small sample of X-ray selected EROs, Stevens et~al. (\cite{stevens})
found similar or even smaller angular radii.
Assuming the minimum redshift computed above, we have derived the intrinsic
dimension for each extended source, listed in Table~\ref{tabzmin}.
The intrinsic effective radii are in the range $3.1<r_{eff}<6.2$ kpc,
similar to that of local $L^*$ galaxies.

In Figure~\ref{fig_re_LK} we show the absolute \ks-band magnitude vs.
intrinsic effective radius for bulge sources, compared with the
relation between luminosity and radius derived from the local
sample of Pahre, Djorgovski, \& de Carvalho (\cite{pahre})
in the near-IR (Angeretti et~al. \cite{rosso}). 
Our sources do not follow the local luminosity-radius relation, most
of them (5/7) being more luminous or smaller than observed locally at a given
radius or luminosity, respectively. A decrease of the intrinsic galaxy 
dimensions at high-$z$ has been suggested also by Trujillo et~al.
(\cite{truj}).
On the other hand, in Fig.\ref{fig_re_LK} we show the same local
relation shifted at $z=1, 1.5$ assuming the luminosity brightening
derived in the K20 survey by Pozzetti et~al. (\cite{K20_5})
with no change in the light profiles. Such a luminosity brightening
could explain the observed radii for the sources at $z\ge1.3$.
The discrepancy between the observed and expected luminosity
for the two objects at the lowest redshifts suggests that
these sources have a higher redshift than the adopted (minimum) one.
Indeed, assuming higher-$z$ would move the sources in
Fig.\ref{fig_re_LK} essentially vertically toward brighter luminosities,  
because of the small redshift dependence of the intrinsic size at $z>1$. 

Concluding, the analysis of the luminosity-radius relation is consistent
with the color-based estimates for the minimum redshift of
our high X/O, hard X-ray selected sources.

%
\subsection{X-ray properties of high X/O EROs}
%
\par\noindent

The X-ray spectra of all ten high X/O sources have been 
fitted with a single power law model plus absorption 
in the source rest--frame assuming for all of them the 
``minimum'' redshift estimated in the previous section.
The quality of the X-ray spectra in terms of S/N ratio is not such to 
simultaneously constrain both the slope and column density.
For this reason the former was fixed at $\Gamma$=1.9,
the average value obtained by fitting the ``bright'' part of the 
\hel \ sample (see Perola et al. (in preparation) for a detailed 
description of the spectral analysis procedures).
The best-fit values for the intrinsic column densities and 
unabsorbed X-ray luminosities are listed in Table~\ref{tabzmin}.
All of the sources in our sample have a best-fit N$_H$ larger than 
$10^{22}$ cm$^{-2}$, and for half of them also the 90\% lower limits
are consistent with significant X-ray absorption.

For three objects in our sample (the two point-like objects and the
disky galaxy) we estimated both the optical reddening and the
X-ray absorbing column density; we calculated the $E_{B-V}/N_{\rm H}$ ratio
and we found it significantly lower than the Galactic standard value
($1.7\times10^{-22}$ mag~cm$^2$, Bohlin, Savage \& Drake \cite{EBVNH}).
The $E_{B-V}/N_{\rm H}$ is lower by a factor of $\sim10$ for the point-like
sources {\tt PKS0312$-$77~\#045} and {\tt Abell2690~\#029A}, whereas
for the disky galaxy the ratio is at least three times lower than the Galactic
one. This confirms the results for X-ray selected AGN (see Maiolino et~al.
\cite{maio01} and references therein), for which there is increasing 
evidence that the AGN circumnuclear region has different dust properties
than the Galactic diffuse interstellar medium.

The unabsorbed 2--10 keV X-ray luminosities of all but two of the
objects in the \hel \ 
are larger than  10$^{44}$ erg s$^{-1}$.  
The presence of more than one object with ERO colors around 
the most probable counterpart suggests that the X--ray emission 
could be due to a cluster of galaxies which remains unresolved with 
\xmm.  While the observed X--ray luminosities would be
consistent with this possibility (Rosati, Borgani \& Norman \cite{RBN}),
the observed spectra are much harder than expected from the
intracluster medium thermal emission at the estimated redshifts. Moreover,
the {\it Chandra} X--ray fluxes, when available, are consistent with the
XMM ones, indicating that these X-ray sources are probably point-like.
The majority of our relatively bright hard X-ray selected sources 
with high X/O and ERO colors are therefore obscured, high-luminosity AGN, 
i.e., Type~II quasars (see also Fiore et~al. \cite{H2X1dF}) .

   \begin{figure}
   \centering
   \includegraphics[width=0.49\textwidth]{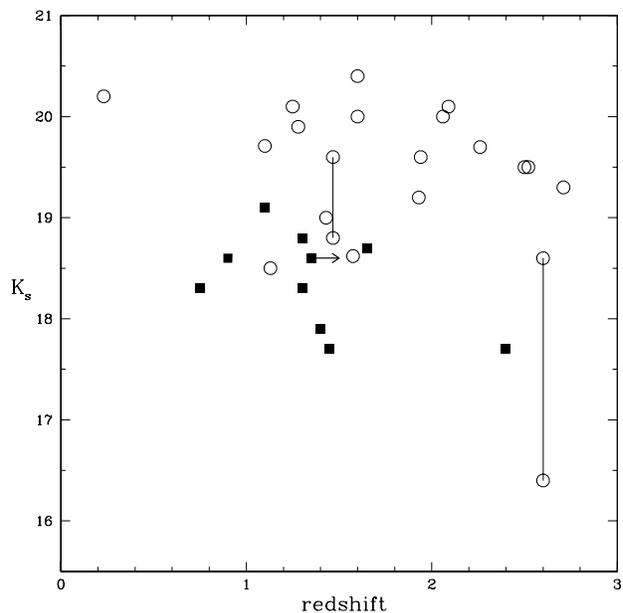}
   \caption{Redshift vs. \ks \ magnitude for our targets (filled square) and 
   for a comparison sample of high X/O sources with ERO colors extracted from
   the literature (empty circles; see text).
   The two pairs of connected circles represent two lensed objects in the
   Abell2390 field (Crawford et~al. \cite{crawford}) with observed and 
   de-magnified \ks.}
    \label{fig_z_K}
    \end{figure}


In order to check whether the X-ray absorption is common among this type of 
objects we have built a comparison sample of high X/O sources
with ERO colors serendipitously discovered at X-ray fluxes
comparable to those of our sample. 
We have considered only those X-ray sources for which a reliable 
spectroscopic or photometric identification was available in the literature:
\#19 and \#25 in the Lockman Hole (Mainieri et~al. \cite{mainieri};
see also Stevens et~al. \cite{stevens}); 
A15 and A18 in A2390 field (Crawford et~al. \cite{crawford}, 
Gandhi, Crawford \& Fabian \cite{gandhi}); N2\_21 and N2\_28 in the
{\tt ELAIS} survey (Willott et~al. \cite{Elais_S1}).
All but one (N2\_21) of the objects in the comparison sample have 
column densities in excess of $10^{22}$ cm$^{-2}$, in agreement 
with the results obtained for the \hel \ sample.

At fainter X-ray fluxes, sources with extremely high X-ray-to-optical
flux ratios have been also discovered in the deep pencil-beam
surveys. From the Barger et~al. (\cite{barger03}) catalog of  
the 2~Ms CDF-N observation, we have selected all of the objects with
X/O$>$10 and \rk $>$5; among these, 13 sources have a spectroscopic or
photometric redshift.
For the two brightest sources absorption column densities in excess
of 10$^{23}$ cm$^{-2}$ were derived through proper spectral analysis 
(see Civano et~al. in prep.); in all but one of 
the remaining cases, the intrinsic column densities inferred from the 
Hardness-Ratio technique are in excess of 10$^{22}$ cm$^{-2}$ (see also 
Alexander et~al. (\cite{alex02}) for the 1~Ms CDF-N observation).
Figure~\ref{fig_z_K} shows that all but one
of the 19 sources in the comparison samples
lie in the same redshift range derived for \hel \ sources, 
thus reinforcing the reliability of our method for the redshift 
determination. 

Deep {\tt HST ACS} observations of
a small sample of 7 {\it Chandra} sources in the CDF-S with extremely
high X/O and red colors (EXOs: Extreme X-ray-to-Optical ratio sources)
have been recently discussed by Koekemoer et~al. (\cite{koeke03}).
None of them was detected down to very faint optical 
magnitudes ($z(850)\simeq$ 28), although these sources
are clearly detected in the \ks~band ($K_{vega} \simeq$ 20.4--23).
It has been proposed that these sources could be 
high-redshift ($z > 6$) L$^{*}$~galaxies powered by a moderately luminous AGN 
(L$_{\rm X}=2-6\times10^{44}$ erg s$^{-1}$).
If they were at lower redshifts, they would be unusually underluminous
galaxies.
The {\it Chandra} EXOs could then constitute the high--redshift tail 
of a population of obscured AGN which is sampled at lower 
redshifts by the \hel \ high X/O sources.
The relatively soft X-ray spectra of {\it Chandra} EXOs, as inferred from
their Hardness-Ratio distribution, would also be consistent with a
high-$z$ hypothesis.

%
%
\section{Summary and Conclusion}
%

The most important results obtained from the analysis of high quality
near--infrared images of hard X-ray sources with X/O $>$ 10 can 
be summarized as follows:

\begin{enumerate}
 \item We have identified 10 out of the 11 \hel \ sources observed with
       {\tt ISAAC} at {\tt VLT} with a relatively bright near infrared
       (\ks$<$19.2) counterpart.

 \item All of them fall in the class of Extremely Red Objects (\rk $>5$)
       and 9 out of 10 satisfy the more stringent criterion \rk $>5.3$. 
       The trend of finding redder sources at fainter optical magnitudes, 
       as well as an increasing fraction of EROs, are well established
       results by various X-ray surveys. However, the three brightest
       counterparts, with \ks$<18$ \emph{and} \rk $>6$, are very rare
       objects also with respect to the galaxy field population and, to
       our knowledge, objects like these were never found by any other
       X-ray survey. 

 \item We found an ERO overdensity around the \hel \ source BPM16274~\#181.
       Five near-infrared sources with \rk $>5$ (plus another bluer object) 
       lie within a radius of about 6\arcsec \ from the hard X-ray source,
       suggesting the presence of a possible high-$z$ cluster/group of galaxies.
 
 \item Thanks to the excellent observing conditions 
       it was possible to perform an accurate analysis of the surface 
       brightness profiles and provide a morphological classification.
       Only two objects are classified as point--like sources, while 
       all of the others are extended, without any evidence of a central
       unresolved nuclear source. 
       The morphological breakdown is dominated by elliptical profiles;
       only in one case the fit with an exponential law (disky galaxy)
       provides a better description of the data.
     
  \item We provide a fairly robust estimate of the minimum redshift for
       each source by comparing the color and morphological information 
       with that available in the literature for spectroscopically
       identified sources and with models for the spectral energy 
       distribution. The \rk \ color of the two point--like objects
       is well matched by a quasar template reddened by dust at
       $z \simeq $ 2. The minimum redshift for extended sources is in
       the range $z$=0.80--1.45.

 \item Most of the sources in the \hel \ sample show substantial intrinsic 
       absorption in their X-ray spectra. A detailed comparison with 
       X-ray selected sources in various deep and medium--deep surveys 
       indicates that heavy obscuration is almost ubiquitous among 
       objects with high X/O and extremely red colors. All but one of
       the sources have unabsorbed X-ray luminosities
       L$_{2-10 {\rm keV}}\ge 10^{44}$~erg~s$^{-1}$, and therefore belong
       to the Type~II quasar population. 
       We wish to stress that the seven elliptical galaxies, with no
       evidence of a central spike, represent, in the whole Hellas2-XMM 1df
       sample, around 10\% of the 72 sources with L$_x\ge
       10^{44}$~erg~s$^{-1}$. Thus, a non-negligible
       fraction of the most luminous AGN, although not heavily obscured 
       (i.e. Compton-thick) in the hard X-ray band, would be missed
       in UV/optical as well as in near-infrared quasar surveys (see also
       Gandhi et~al. \cite{g03}).
      
   \end{enumerate}
  
 Finally, we wish to stress that near--infrared observations of hard X-ray
 sources selected on the basis of a high X/O ratio have proven to be a powerful
 technique aimed at studying the hosts of high-z Type~II AGN, whose obscured
 nuclei do not affect the host galaxy morphologies. Moreover, our results do
 indicate that X-ray surveys constitute an efficient mean to select
 high-redshift, massive elliptical galaxies and that the ``obscured'' nuclear
 activity takes place within otherwise normal galaxies,
 which already formed the bulk of their stars at high redshift.
 Deep near--infrared spectroscopy of these
 sources would allow to test this possibility further.

%
%
\begin{acknowledgements}
 We are grateful to G.~Zamorani for insightful discussions and to R.~Sancisi 
 e M.G.~Stirpe for their careful reading of the paper.
 We thanks the K20 team for making their spectroscopic results available
 before publication.
 This work was partially supported by the Italian Space Agency 
 (ASI) under grants I/R/073/01 and I/R/057/02 and by INAF (grant \#270/2003). 
\end{acknowledgements}
%
%

%
%
\end{document}